# Revisiting Synthesis Model in Sparse Audio Declipper

Pavel Záviška[1], Pavel Rajmic[1], Zdeněk Průša[2], and Vítězslav Veselý[3]

[1]Signal Processing Laboratory, Brno University of Technology, Czech Republic
rajmic@vutbr.cz
[2]Acoustics Research Institute, Austrian Academy of Sciences, Vienna, Austria
zdenek.prusa@oeaw.ac.at
[3]Faculty of Mechanical Engineering, Brno University of Technology, Czech Republic

**Abstract.** The state of the art in audio declipping has currently been achieved by SPADE (SParse Audio DEclipper) algorithm by Kitić et al. Until now, the synthesis/sparse variant, S-SPADE, has been considered significantly slower than its analysis/cosparse counterpart, A-SPADE. It turns out that the opposite is true: by exploiting a recent projection lemma, individual iterations of both algorithms can be made equally computationally expensive, while S-SPADE tends to require considerably fewer iterations to converge. In this paper, the two algorithms are compared across a range of parameters such as the window length, window overlap and redundancy of the transform. The experiments show that although S-SPADE typically converges faster, the average performance in terms of restoration quality is not superior to A-SPADE.

## 1 Introduction

Clipping is a non-linear form of signal distortion which appears in the context of signal acquisition, processing or transmission. In general, clipping occurs when the signal amplitude gets outside of the allowed dynamic range. Along with missing samples and additive noise, clipping is one of the most common types of audio signal degradation. Not only does clipping have a negative effect on perceived audio quality [35], it also degrades the accuracy of automatic speech recognition [25,19,34]. This motivates a restoration task usually termed *declipping*, i.e. the recovery of signal samples that originally lay outside the recognized range.

In this work, we concentrate on the case of the so-called *hard clip* degradation, where the waveform of the signal is simply truncated such that the signal value cannot leave the interval $[-\theta_c, \theta_c]$. If the vector $\mathbf{x} \in \mathbb{R}^N$ denotes the original discrete-time signal, then the respective hard-clipped signal is

$$\mathbf{y}[n] = \begin{cases} \mathbf{x}[n] & \text{for } |\mathbf{x}[n]| < \theta_c, \\ \theta_c \cdot \text{sgn}(\mathbf{x}[n]) & \text{for } |\mathbf{x}[n]| \geq \theta_c, \end{cases} \quad (1)$$

i.e. hard clipping acts elementwise, wasting information in the peaks of **x** that exceed the *clipping threshold* $\theta_\mathrm{c}$.

In the past, several attempts were made to perform declipping. Since declipping is inherently ill-posed, any method attacking the problem must introduce an assumption about the signal. As a short review of the field, we mention a method based on autoregressive signal modelling [21], a method based on the knowledge of the original signal bandwidth [1], statistical approaches [17,15], and simple, even if not quite effective algorithms in [11], [32] and [26]. The quality of restoration was significantly elevated when models involving the *sparsity* of the audio signal were introduced. In such models, it is assumed that there is a transform which either approximates the signal well using a low number of nonzero coefficients (the synthesis/sparse model), or the transform applied to the signal produces a low number of nonzero coefficients (the analysis/cosparse model) [8,13,27]. Suitable transform are usually time-frequency operators such as the Discrete Fourier Transform (DFT), the Discrete Cosine Transform (DCT), or the Discrete Gabor Transform (DGT), also known as the Short-time Fourier Transform (STFT) [18,9,14].

The very first method for sparse declipping was published in [2]; it was based on the greedy approximation of a signal within the reliable (i.e. not clipped) parts. Many alternative approaches appeared after this successful paper, such as in [36] that brought convex optimization into play, or in [33] where the authors forced a structure into the sparse coefficients (known as "structured" or "social" sparsity). Article [12] shows that the introduction of a psychoacoustic masking model (although very simple) improves the perceived quality of the restored signal. Besides [6], which relies on non-negative matrix factorization, all the mentioned papers process the signal from the synthesis viewpoint. More recently, a series of papers have considered the declipping problem from the analysis side as well [22,23,24], while [24] is considered a state-of-the-art declipper.

In this paper, we show that using a novel projection lemma we were able to derive a synthesis-based algorithm which is even faster than the analysis-based algorithm in [24]. Our experiments show that our algorithm, nevertheless, does not outperform the analysis version in terms of quality of restoration.

In Sec. 2, the declipping problem is formalized. Then in Sec. 3, the two versions of the SPADE algorithm [24] are reviewed, and the new projection lemma is exploited to develop a fast synthesis-based algorithm. Sec. 4 reports on experiments that have been run.

## 2  Problem Formulation

Assume that a signal $\mathbf{x} \in \mathbb{R}^N$ has been clipped according to (1). We observe the clipped signal $\mathbf{y} \in \mathbb{R}^N$. We suppose that it is possible to divide the signal samples into three sets $R$, $H$ and $L$, which correspond to "reliable" samples and samples that have been clipped to the "high" and "low" clipping thresholds, respectively. To select only samples of a specific set, linear *restriction operators* $M_\text{R}$, $M_\text{H}$ and $M_\text{L}$ will be used. Note that if these sets are not known in advance, they can be trivially induced from the particular values of $\mathbf{y}$.

Denote the declipped signal by $\hat{\mathbf{x}}$. While performing any declipping algorithm, it is natural to enforce that the samples $M_\text{R}\hat{\mathbf{x}}$ match the reliable samples $M_\text{R}\mathbf{y}$. The authors of the C-IHT algorithm [22] call this approach *consistent*. Our approach obeys full *consistency*, meaning that in addition, the samples $M_\text{H}\hat{\mathbf{x}}$ should lie at or above $\theta_\text{c}$ and the samples from $M_\text{L}\hat{\mathbf{x}}$ should not lie above $-\theta_\text{c}$. These requirements are formalized by defining a set of signals $\Gamma$, consistent with the three conditions:

$$\Gamma = \Gamma(\mathbf{y}) = \{\hat{\mathbf{x}} \mid M_\text{R}\hat{\mathbf{x}} = M_\text{R}\mathbf{y}, M_\text{H}\hat{\mathbf{x}} \geq \theta_\text{c}, M_\text{L}\hat{\mathbf{x}} \leq -\theta_\text{c}\}. \qquad (2)$$

In line with the recent literature, the fact that many musical signals are sparse with respect to a (time-)frequency transform will be exploited. To put it in words, one would like to find signal $\hat{\mathbf{x}}$ that is the most sparse among all signals belonging to the consistency set $\Gamma$. The state of the art declipping results are achieved by the SPADE algorithm, which will be described in the next section. It comes in two variants, based on either the synthesis (sparse) or the analysis (cosparse) understanding of "sparsity" [27].

## 3 The SPADE algorithm

SPADE (SParse Audio DEclipper) [24] is a heuristic declipping algorithm, approximating the solution of the following non-convex, NP-hard synthesis- or analysis-regularized inverse problems:

$$\min_{\mathbf{x},\mathbf{z}} \|\mathbf{z}\|_0 \quad \text{s.t.} \quad \mathbf{x} \in \Gamma(\mathbf{y}) \text{ and } \|\mathbf{x} - \mathbf{D}\mathbf{z}\|_2 \leq \epsilon, \qquad (3)$$

$$\min_{\mathbf{x},\mathbf{z}} \|\mathbf{z}\|_0 \quad \text{s.t.} \quad \mathbf{x} \in \Gamma(\mathbf{y}) \text{ and } \|\mathbf{A}\mathbf{x} - \mathbf{z}\|_2 \leq \epsilon. \qquad (4)$$

Here $\|\mathbf{z}\|_0$ is the $\ell_0$ pseudonorm measuring the sparsity, i.e. counting the nonzero elements of $\mathbf{z}$. The $\ell_2$ constraint delimits the distance between the estimate and its sparse approximation. The linear operator $\mathbf{D} : \mathbb{C}^P \mapsto \mathbb{R}^N$ is the synthesis operator, with $N \leq P$; if regarded as a matrix in (3), it takes coefficients $\mathbf{z}$ and forms the signal as the linear combination of its columns. Matrix $\mathbf{D}$ is often called the *dictionary* [8]. In (4), the analysis operator $\mathbf{A} : \mathbb{R}^N \mapsto \mathbb{C}^P$ is considered, which analyses the signal and produces its transform coefficients. In order to be able to compare the two approaches, we naturally restrict ourselves to the case when the operators are mutually adjoint, $\mathbf{A} = \mathbf{D}^*$.

Note that problems (3) and (4) both seek the signal and its coefficients simultaneously and that both of them fall into a common, recently introduced general signal restoration framework, see [16]. Both $\mathbf{A}$ and $\mathbf{D}$ are assumed full rank, $N$, and both formulations produce equal results when $\mathbf{D}$ is a unitary operator $\mathbf{A} = \mathbf{D}^{-1}$ (the same will hold for the approximate solutions by SPADE).

It should be noted that in SPADE, the above optimization problems are solved frame-by-frame, i.e. the signal is segmented into possibly overlapping time chunks, and windowed. Problems (3) and (4) are then solved individually on each such segment, and the output is formed using a common overlap-add procedure. This allows real-time processing, and at the same time the time-frequency structure of the processing is preserved. Specifically, the windowed (I)DFT is used in place of the operators $\mathbf{A}$ and $\mathbf{D}$, possibly with frequency oversampling [9].

SPADE addresses the above two problems by a modified ADMM algorithm [5,7] resulting in the synthesis SPADE (S-SPADE) as shown in Alg. 1, and the analysis SPADE (A-SPADE) as given in Alg. 2.

| **Algorithm 1:** S-SPADE | **Algorithm 2:** A-SPADE |
|---|---|
| **Require:** $\mathbf{D}, \mathbf{y}, M_\mathrm{R}, M_\mathrm{H}, M_\mathrm{L}, s, r, \epsilon$ | **Require:** $\mathbf{A}, \mathbf{y}, M_\mathrm{R}, M_\mathrm{H}, M_\mathrm{L}, s, r, \epsilon$ |
| 1  $\hat{\mathbf{z}}^{(0)} = \mathbf{D}^*\mathbf{y}, \mathbf{u}^{(0)} = \mathbf{0}, i=1, k=s$ | 1  $\hat{\mathbf{x}}^{(0)} = \mathbf{y}, \mathbf{u}^{(0)} = \mathbf{0}, i=1, k=s$ |
| 2  $\bar{\mathbf{z}}^{(i)} = \mathcal{H}_k\left(\hat{\mathbf{z}}^{(i-1)} + \mathbf{u}^{(i-1)}\right)$ | 2  $\bar{\mathbf{z}}^{(i)} = \mathcal{H}_k\left(\mathbf{A}\hat{\mathbf{x}}^{(i-1)} + \mathbf{u}^{(i-1)}\right)$ |
| 3  $\hat{\mathbf{z}}^{(i)} = \arg\min_\mathbf{z} \|\mathbf{z} - \bar{\mathbf{z}}^{(i)} + \mathbf{u}^{(i-1)}\|_2^2$ s.t. $\mathbf{Dz} \in \Gamma$ | 3  $\hat{\mathbf{x}}^{(i)} = \arg\min_\mathbf{x} \|\mathbf{Ax} - \bar{\mathbf{z}}^{(i)} + \mathbf{u}^{(i-1)}\|_2^2$ s.t. $\mathbf{x} \in \Gamma$ |
| 4  **if** $\|\hat{\mathbf{z}}^{(i)} - \bar{\mathbf{z}}^{(i)}\|_2 \le \epsilon$ **then** | 4  **if** $\|\mathbf{A}\hat{\mathbf{x}}^{(i)} - \bar{\mathbf{z}}^{(i)}\|_2 \le \epsilon$ **then** |
| 5     terminate | 5     terminate |
| 6  **else** | 6  **else** |
| 7     $\mathbf{u}^{(i)} = \mathbf{u}^{(i-1)} + \hat{\mathbf{z}}^{(i)} - \bar{\mathbf{z}}^{(i)}$ | 7     $\mathbf{u}^{(i)} = \mathbf{u}^{(i-1)} + \mathbf{A}\hat{\mathbf{x}}^{(i)} - \bar{\mathbf{z}}^{(i)}$ |
| 8     $i \leftarrow i+1$ | 8     $i \leftarrow i+1$ |
| 9     **if** $i \bmod r = 0$ **then** | 9     **if** $i \bmod r = 0$ **then** |
| 10        $k \leftarrow k+s$ | 10        $k \leftarrow k+s$ |
| 11    **end** | 11    **end** |
| 12    go to 2 | 12    go to 2 |
| 13 **end** | 13 **end** |
| 14 **return** $\hat{\mathbf{x}} = \mathbf{D}\hat{\mathbf{z}}^{(i)}$ | 14 **return** $\hat{\mathbf{x}} = \hat{\mathbf{x}}^{(i)}$ |

Both SPADE algorithms rely on two principal steps. The first of them is the hard thresholding $\mathcal{H}_k$. This operator enforces sparsity by setting all but $k$ largest components of the input vector to zero. In practice, the sparsity $k$ of signals is unknown, therefore SPADE performs *sparsity relaxation*: in every $r$-th iteration the variable $k$ is incremented by $s$ until the constraint embodied by the $\ell_2$ norm is smaller than $\epsilon$. The second main step is the projection onto $\Gamma$ in order to keep the consistency given by (2) and will be discussed in the following.

### 3.1 Projection in A-SPADE

The projection in SPADE (row 3 in both Algorithms 2 and 1) constitutes the computationally most demanding step. For general $\mathbf{A}$ and $\mathbf{D}$, such projections are achievable only via iterative algorithms.

The projection in A-SPADE is written as an optimization problem where one has to find a consistent signal $\mathbf{x}$ such that its analysis coefficients $\mathbf{Ax}$ are the closest possible with respect to the given $(\bar{\mathbf{z}}^{(i)} - \mathbf{u}^{(i-1)})$. The authors of [24] exploit the advantage that when $\mathbf{A}^*$ is a tight Parseval frame, i.e. $\mathbf{A}^*\mathbf{A} = \mathbf{DD}^* = \mathbf{DA}$ are all identity operators [9], then the projection can be done elementwise in the time domain, such that

$$\hat{\mathbf{x}}^{(i)} = \text{proj}_\Gamma \left( \mathbf{A}^*(\bar{\mathbf{z}}^{(i)} - \mathbf{u}^{(i-1)}) \right) \tag{5}$$

where $\text{proj}_\Gamma$ is the operator of orthogonal projection onto a convex set $\Gamma$, in our case defined as

$$[\text{proj}_\Gamma(\mathbf{w})]_n = \begin{cases} [\mathbf{y}]_n & \text{for } n \in R, \\ \max\{[\mathbf{w}]_n, \theta_c\} & \text{for } n \in H, \\ \min\{[\mathbf{w}]_n, -\theta_c\} & \text{for } n \in L, \end{cases} \tag{6}$$

where $[\cdot]_n$ denotes the $n$-th element of a vector.

We now rewrite the projection into a more convenient form. Let $\tilde{\mathbb{R}}$ denote the extended real line, i.e. $\tilde{\mathbb{R}} = \mathbb{R} \cup \{-\infty, \infty\}$. Define the lower and upper bounding vectors $\mathbf{b}_\text{L}, \mathbf{b}_\text{H} \in \tilde{\mathbb{R}}$ such that

$$[\mathbf{b}_\text{L}]_n = \begin{cases} [\mathbf{y}]_n & \text{for } n \in R, \\ \theta_c & \text{for } n \in H, \\ -\infty & \text{for } n \in L, \end{cases} \quad [\mathbf{b}_\text{H}]_n = \begin{cases} [\mathbf{y}]_n & \text{for } n \in R, \\ \infty & \text{for } n \in H, \\ -\theta_c & \text{for } n \in L. \end{cases} \tag{7}$$

Recognizing that the multidimensional interval $[\mathbf{b}_\text{L}, \mathbf{b}_\text{H}]$ matches the set of feasible solutions (2), specifically $\Gamma = \{\mathbf{x} \mid \mathbf{b}_\text{L} \leq \mathbf{x} \leq \mathbf{b}_\text{H}\}$, the final A-SPADE projection formula (5) can be written as

$$\hat{\mathbf{x}}^{(i)} = \text{proj}_{[\mathbf{b}_\text{L}, \mathbf{b}_\text{H}]} (\mathbf{A}^* \mathbf{v}) \quad \text{with} \quad \mathbf{v} = \bar{\mathbf{z}}^{(i)} - \mathbf{u}^{(i-1)}. \tag{8}$$

The projection onto the interval can be implemented as

$$\text{proj}_{[\mathbf{b}_\text{L}, \mathbf{b}_\text{H}]}(\mathbf{w}) = \min\{\max\{\mathbf{b}_\text{L}, \mathbf{w}\}, \mathbf{b}_\text{H}\}, \tag{9}$$

with the min and max functions returning pairwise extremes element by element.

Note that restricting to the Parseval tight frames in applications is not an issue [24,3,4,31,30,28,20].

### 3.2 Projection in S-SPADE

For S-SPADE the situation is different. The projection has to be done in the domain of coefficients. The authors of [24] claim that the projection needs to be computed iteratively and that a somewhat efficient implementation can be achieved with $\mathbf{D}$ forming a tight Parseval frame. Still, [24] reports many times higher computational time for S-SPADE compared to A-SPADE.

We will show that it is possible to use an explicit formula to compute the projection in S-SPADE, making the two algorithms identical from the point of view of complexity per iteration. Our goal is to find the optimizer

$$\hat{\mathbf{z}}^{(i)} = \arg\min_{\mathbf{z}} \|(\bar{\mathbf{z}}^{(i)} - \mathbf{u}^{(i-1)}) - \mathbf{z}\|_2^2 \quad \text{s.t.} \quad \mathbf{Dz} \in \Gamma. \tag{10}$$

The following lemma can be found in several variations, see, for example, [29] or [10]. We introduce a real-setting version for simplicity.

**Lemma:** *Let the operator $\mathbf{D} : \mathbb{R}^P \mapsto \mathbb{R}^N$, $N \leq P$, full-rank, $\mathbf{DD}^\top$ identity. Let the multidimensional interval bounds $\mathbf{b}_\mathrm{L}, \mathbf{b}_\mathrm{H} \in \tilde{\mathbb{R}}^N$, $\mathbf{b}_\mathrm{L} \leq \mathbf{b}_\mathrm{H}$. Then the projection of a vector $\mathbf{v} \in \mathbb{R}^N$, respectively denoted and defined by*

$$\mathrm{proj}_{\{\mathbf{x} \,|\, \mathbf{Dx} \in [\mathbf{b}_\mathrm{L}, \mathbf{b}_\mathrm{H}]\}}(\mathbf{v}) := \arg\min_{\mathbf{u}} \|\mathbf{v} - \mathbf{u}\|_2 \; s.t. \; \mathbf{Du} \in [\mathbf{b}_\mathrm{L}, \mathbf{b}_\mathrm{H}],$$

*can be evaluated as*

$$\mathrm{proj}_{\{\mathbf{x} \,|\, \mathbf{Dx} \in [\mathbf{b}_\mathrm{L}, \mathbf{b}_\mathrm{H}]\}}(\mathbf{v}) = \mathbf{v} - \mathbf{D}^\top \left( \mathbf{Dv} - \mathrm{proj}_{[\mathbf{b}_\mathrm{L}, \mathbf{b}_\mathrm{H}]}(\mathbf{Dv}) \right). \tag{11}$$

In our application, we will need a complex $\mathbf{D}$ with a special (time-)frequency structure. Indeed, our $\mathbf{D}$ will be the synthesis operator of (possibly redundant) discrete Fourier and Gabor tight frames [9]. In such cases, it is only necessary to substitute $\mathbf{D}^\top$ by $\mathbf{D}^*$ in (11). The proof of such an extended lemma, however, gets much more involved by switching to the complex case, and therefore we omit it for simplicity of presentation, as we plan to publish it in a separate paper (currently in preparation).

Using $\mathbf{b}_\mathrm{L}$ and $\mathbf{b}_\mathrm{H}$ as defined above, the projection (10) can be written as

$$\hat{\mathbf{z}}^{(i)} = \mathbf{v} - \mathbf{D}^* \left( \mathbf{Dv} - \mathrm{proj}_{[\mathbf{b}_\mathrm{L}, \mathbf{b}_\mathrm{H}]}(\mathbf{Dv}) \right) \quad \text{with} \quad \mathbf{v} = \bar{\mathbf{z}}^{(i)} - \mathbf{u}^{(i-1)}. \tag{12}$$

### 3.3  Comparing computational complexity

In both SPADE algorithms, the computational cost is dominated by the analysis and synthesis operators. Returning to Algorithms 1 and 2, we see that A-SPADE requires one analysis in step 2 and one synthesis in the projection (5). In the case of S-SPADE, the projection is the only demanding calculation, and according to the new formula (12), it requires one synthesis and one analysis. This shows that *per iteration*, the two algorithms are equally demanding. This breaks down the disadvantage of S-SPADE as presented in [24].

## 4  Experiments and results

Experiments are designed to compare A-SPADE and S-SPADE algorithms in terms of quality of restoration and computational time. The quality of restoration is evaluated using $\Delta\mathrm{SDR}$, which expresses the signal-to-distortion ratio improvement, according to the following formula:

$$\Delta\mathrm{SDR} = \mathrm{SDR}(\mathbf{x}, \hat{\mathbf{x}}) - \mathrm{SDR}(\mathbf{x}, \mathbf{y}) \tag{13}$$

where $\mathbf{x}$ represents the original signal (known in our study), $\mathbf{y}$ is the clipped signal and $\hat{\mathbf{x}}$ is the reconstructed signal, while the SDR itself is defined as

$$\mathrm{SDR}(\mathbf{u}, \mathbf{v}) = 10 \log_{10} \frac{\|\mathbf{u}\|_2^2}{\|\mathbf{u} - \mathbf{v}\|_2^2} \; [\mathrm{dB}]. \tag{14}$$

The results below are usually presented in *average* $\Delta$SDR values, taking the arithmetic mean of the particular values from all the tested audio signals in dB.

The advantage of using $\Delta$SDR over the plain SDR is that the $\Delta$SDR value remains the same irrespective of whether the SDR is computed on the whole signal or on the clipped samples only. (This can be easily shown directly from (13), using the fact that our algorithms are consistent, i.e. the reliable samples of the recovered signal and of the clipped signal match.)

Experiments were performed on five audio samples with an approximate duration of 5 seconds at a sampling frequency of 16 kHz. These excerpts were thoroughly selected to be diverse enough in tonal content and in sparsity with respect to the time-frequency transform. As a preprocessing step, the signals under consideration were peak-normalized and then artificially clipped using multiple clipping thresholds, $\theta_c \in \{0.1, 0.2, \ldots, 0.9\}$. The algorithms were implemented in MATLAB R2017a and ran on a PC with Intel i7-3770, 16 GB RAM in single thread mode.

Note that some authors ([22,23,24,16]) evaluate the quality of restoration depending on the *input SDR*, while in this paper we plot the results against the *clipping threshold* $\theta_c$. To get a notion of their relationship, we attach Table 1 which shows both the $\theta_c$ and the average input SDR values.

**Table 1.** Average SDR values for a particular clipping threshold $\theta_c$ computed on the test signals as a whole and on the clipped samples only.

| $\theta_c$ [–] | 0.1 | 0.2 | 0.3 | 0.4 | 0.5 | 0.6 | 0.7 | 0.8 | 0.9 |
|---|---|---|---|---|---|---|---|---|---|
| SDR [dB] whole signal | 3.71 | 7.49 | 11.40 | 15.54 | 20.15 | 25.32 | 31.44 | 38.74 | 48.11 |
| SDR [dB] clipped samples | 3.46 | 6.30 | 8.68 | 10.78 | 13.16 | 15.22 | 18.04 | 20.37 | 23.63 |

Although the original SPADE algorithms [24] were purely designed to process individual windowed time-frames, one after another, we also include an experiment using SPADE on the whole signal, considering the DGT coefficients all at once (Sec. 4.1). Then in Sec. 4.2, the classical SPADE setup is investigated, and in later sections the influence of the window length, transform redundancy and window overlap is considered. Note that in this paper, the term *redundancy* specifies the rate of oversampling in the frequency domain—for example, using an oversampled Fourier analysis with 2048 frequency channels applied to a signal of length 1024 means redundancy 2.

### 4.1   SPADE applied to whole signal

Fig. 1 presents the SDR improvement ($\Delta$SDR) for signals processed with SPADE as a whole. The relaxation parameters of both algorithms are set to $r = 1, s = 100$ and $\epsilon = 0.1$. In this experiment, the most common DGT declipping setting such as 1024-sample-long Hann window and 75% overlap is used, although according to Sec. 4.3 such setting favors the analysis approach, which performs better with shorter windows. Redundancy levels 1, 2 and 4 are achieved by setting the number of frequency channels $M$ to 1024, 2048 and 4096. The black line in Fig. 1 denotes the result of S-SPADE with redundancy 1. However, this is identical to A-SPADE results with the same redundancy—in such a case, $\mathbf{D}^{-1} = \mathbf{A}$ and the algorithms perform equally (see Sec. 3).

An iteration of S-SPADE is typically slightly slower (approximately by 2 %) than an iteration of A-SPADE. However, in general, S-SPADE needs fewer iterations to converge. The algorithm is considered converged if the condition on row 4 in both Algorithms 1 and 2 gets fulfilled, i.e. the termination function falls under a prescribed $\epsilon$. Fig. 2 presents the computation times; it is clear that S-SPADE converges significantly faster than A-SPADE does, especially at higher redundancies. The average course of the termination function is presented in Fig. 7.

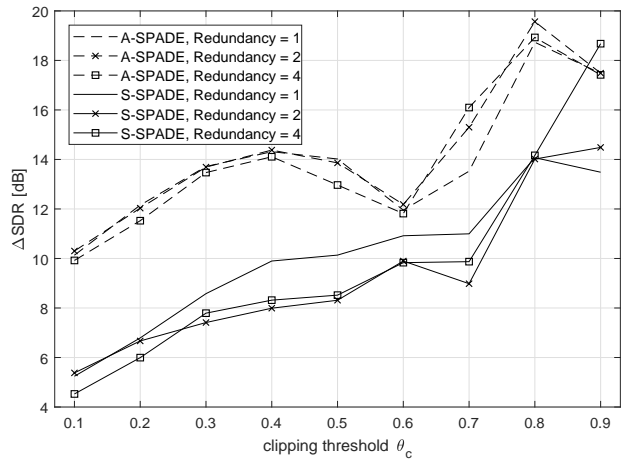

**Fig. 1.** Declipping performance in terms of $\Delta$SDR performed on the whole signal.

### 4.2   SPADE with signal segmentation

The disadvantage of the approach in Sec. 4.1 is that the largest time-frequency coefficients are selected from the whole signal, and the placement of the coefficients over time is not taken into account. This can easily result in selecting a group of significant coefficients from a short time period and ignoring coefficients that are significant rather locally. Thus, (as will be confirmed by experiments) it is more beneficial to process the signal with SPADE *block by block*.

For this experiment, the relaxation parameters are set according to the original paper [24], i.e. $r = 1, s = 1$ and $\epsilon = 0.1$. The transform parameters are set as

Revisiting Synthesis Model in Sparse Audio Declipper    9

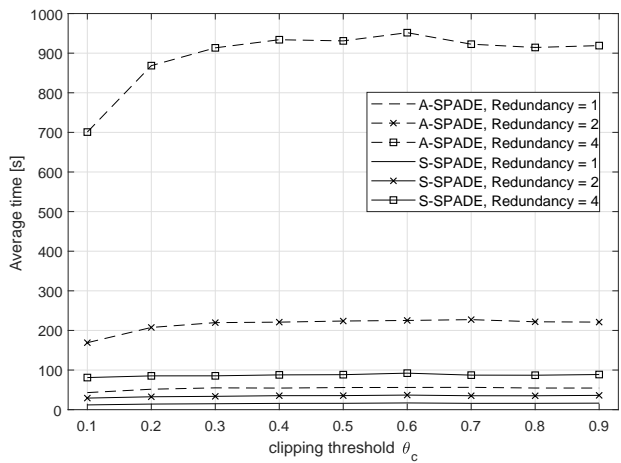

**Fig. 2.** Average computational times for declipping performed on the whole signal.

in the previous experiment, i.e. the sliding Hann window 1024 samples long with 75% overlap and DFT with redundancy 1, 2 and 4 is used in each time-block.

Fig. 3 presents $\Delta$SDR results of both the A-SPADE and S-SPADE algorithms with processing by blocks. Even in this experiment, A-SPADE performs slightly better but it is worth repeating that the choice of the window length suits better the A-SPADE. Interestingly, A-SPADE performs somewhat better with more redundant DFT, while S-SPADE, on the contrary, performs best with no redundancy at all.

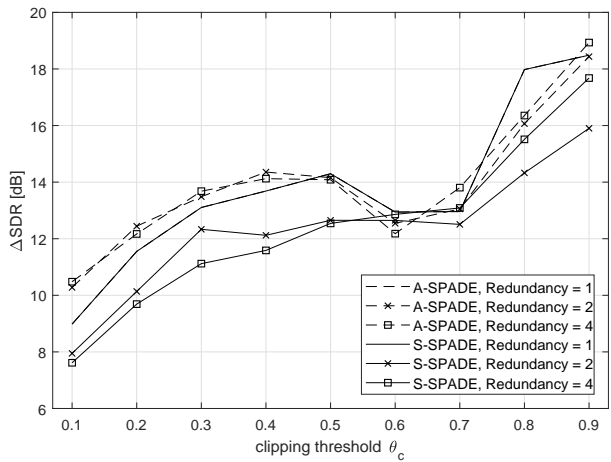

**Fig. 3.** Declipping performance in terms of $\Delta$SDR performed with signal segmentation.

Apart from the overall performance, we also evaluated the two algorithms locally—we wanted to know whether A-SPADE or S-SPADE better recovers the signal *within a short time range*. Figures 4 and 5 demonstrate SDR results on two audio signals using the Hann window 1024 samples long with 75 % overlap and DFT with redundancy 2. For each 2048-sample-long block we computed two corresponding SDR values, which are represented by a marker in the scatter plot. For clarity, we only used clipping thresholds from 0.1 to 0.5. The SDR values were computed using formula (14) on the whole signals; computing SDRs on clipped samples would only reflect in a pure shift of axes in the scatter plot.

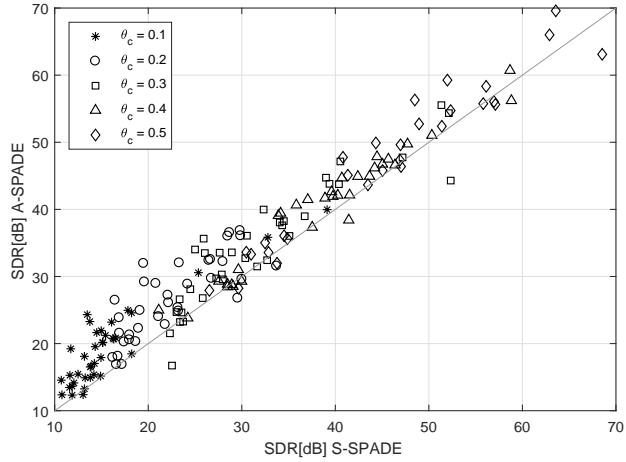

**Fig. 4.** Scatter plot of SDR values for both S-SPADE and A-SPADE computed locally on sliding blocks 2048 samples long. It is clear that in most time chunks, A-SPADE results are better than those of S-SPADE. Results shown here are for the acoustic guitar signal, but nevertheless such a scatter plot is obtained for most of our test signals.

When redundancy 1 is used, the two algorithms perform identically, and they also terminate after the same number of iterations. In light of this, computation times presented in Fig. 6 show that in such a case, A-SPADE is marginally faster. For more redundant transforms, S-SPADE needs fewer iterations to fulfill the termination criterion and its solution is obtained more quickly.

Fig. 7 presents the average course of the termination function (row 4 in both Algorithm 1 and 2). For S-SPADE, this function decreases faster, causing the whole algorithm to converge in fewer iterations.

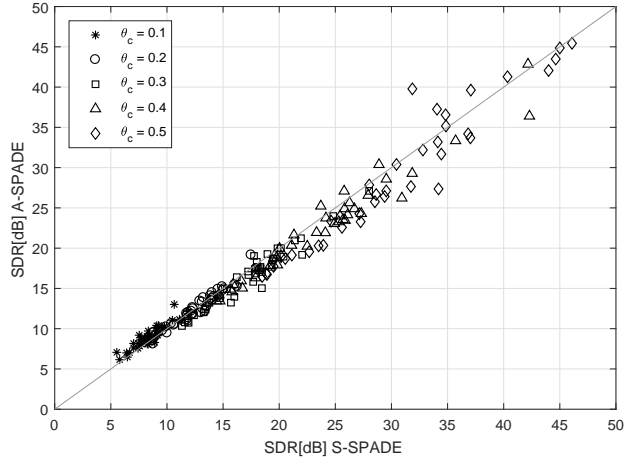

**Fig. 5.** Scatter plot of SDR values for both S-SPADE and A-SPADE computed locally on sliding blocks 2048 samples long. More than half the time chunks resulted in markers below the identity line, indicating that S-SPADE returned better results. The audiosignal used here is a heavy metal song; note that it was hard to find a signal with such a scatter plot. The result may seem optimistic for S-SPADE—however considering that already the input signal has been clipped on purpose (as is commonly done in this music genre, making the signal far from being sparse), it in turn means that the A-SPADE in effect outperforms the S-SPADE again.

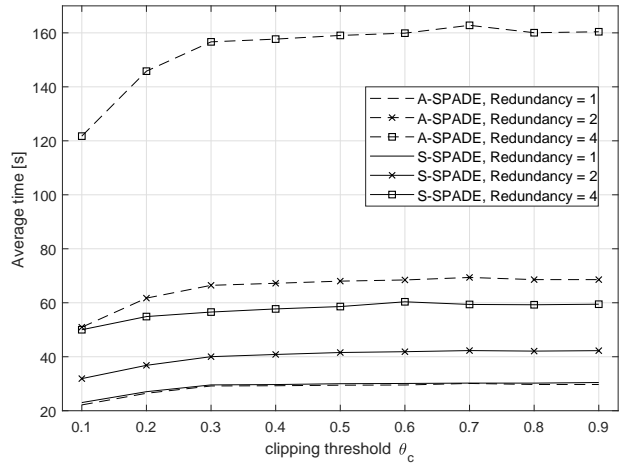

**Fig. 6.** Declipping performance in terms of average computational time performed block-by-block.

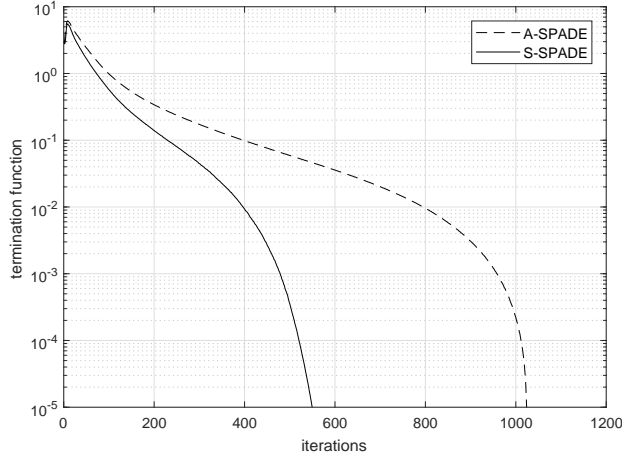

**Fig. 7.** Average course of the termination function during iterations for redundancy 2.

### 4.3 Window length

In many declipping algorithms where processing by blocks or via STFT is done, such as [2,22,33,23], the usual block (or window) length is set to 1024 samples. This experiment is designed to compare both SPADE algorithms depending on selected window length. Signals are processed block by block similarly to the previous experiment, except that the redundancy of the DFT is 2 and the length of the sliding window is set to 512, 1024, 2048 and 4096 samples. In all four cases, the window overlap is fixed to 75 %.

Figues 8 and 9 present $\Delta$SDR results depending on the window length for A-SPADE and S-SPADE respectively. For the analysis approach, the length of 2048 samples seems to give the best results for most clipping thresholds. When using shorter (512 samples) or longer (4096 samples) windows, the SDR performance drops down approximately by 2 dB. On the other hand, according to Fig. 9 the synthesis approach performs better with longer windows. The length of 2048 samples seems to be optimal for S-SPADE as well, but the 4096-sample-long window performs by 2 dB better than the 1024 long one.

As far as the computation time is concerned, a longer window means longer computation time. Average computational times for the window lengths 512, 1024, 2048 and 4096 are listed in Tab. 2.

**Table 2.** Average computation times in seconds depending on window length using overcomplete DFT with redundancy 2.

| window length | 512 | 1024 | 2048 | 4096 |
|---|---|---|---|---|
| A-SPADE | 53 | 68 | 104 | 207 |
| S-SPADE | 34 | 41 | 60 | 115 |

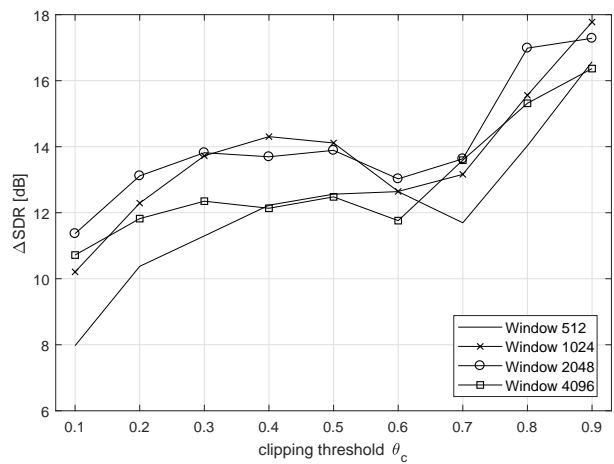

**Fig. 8.** Declipping performance of A-SPADE in terms of ΔSDR for different window lengths.

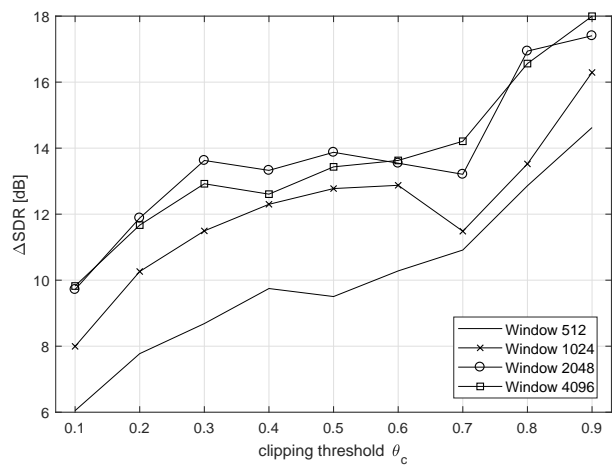

**Fig. 9.** Declipping performance of S-SPADE in terms of ΔSDR for different window lengths.

### 4.4 Window overlap

Window overlap is also an important parameter of the transform; it affects not only the quality of restoration but also the computational time. Therefore, in this experiment, the restoration quality depending on window overlap is explored. As in the previous experiment, DFT with redundancy 2 and Hann window 1024 samples long is used.

Fig. 10 shows an expectable fact that the bigger the overlap is set, the better in terms of SDR the results are produced. In line with the results given above, A-SPADE performs slightly better than S-SPADE due to the chosen window length. More interestingly, the performance of the synthesis version drops significantly when the overlap is set to 25 %. Thus, for a good reconstruction, it is necessary to set the window overlap at least to 50 % of the window length.

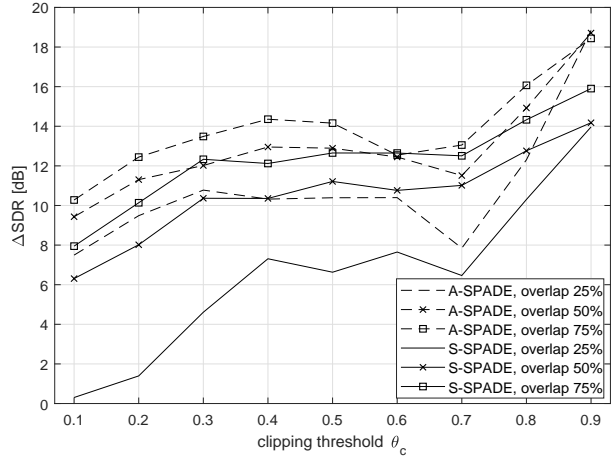

**Fig. 10.** Declipping performance in terms of $\Delta$SDR for different window overlaps.

Table 3 shows the average computation times for overlaps of 25, 50 and 75 % of the window length. We note that an overlap larger than 75 % further increases the computation time but does not bring much improvement in terms of SDR (these facts are not shown).

**Table 3.** Average computation times in seconds depending on window overlap.

| overlap | 25 % | 50 % | 75 % |
|---|---|---|---|
| A-SPADE | 22 | 34 | 68 |
| S-SPADE | 14 | 21 | 41 |

## 5  Implementation

Developing the idea of reproducible research, we make our Matlab codes publicly available. The bundle is downloadable from URL
`http://www.utko.feec.vutbr.cz/~rajmic/software/aspade_vs_sspade.zip`
The main file of the SPADE package is a batch file `declipping_main.m`, reading the audio, normalizing and clipping the signal by calling `hard_clip.m`. It is possible to set transform parameters, such as the window length, overlap, window type, and the transform redundancy.

To process signals block-by-block, `spade_segmentation.m` is used. This function performs signal padding, dividing into blocks, multiplying by the analysis window and, after processing, multiplying by the synthesis window and folding blocks back together (in the common "overlap-add" manner). The SPADE algorithm itself is implemented in two m-files: `aspade.m` for the analysis version and `sspade.m` for the synthesis version.

Recall that the spectrum of a real signal is provided with the complex-conjugate structure. Hard thresholding, performed by `hard_thresholding.m`, takes therefore the oversampled spectrum and thresholds the respective *pairs* of complex entries, in order to keep the signal real. Projections onto the set of feasible solutions are implemented in two m-files. Projection in the time domain for A-SPADE according to (6) is implemented in `proj_time.m`. S-SPADE uses `proj_parse_frame.m` according to (11).

## 6  Conclusion

We exploited a novel projection lemma to speed up the synthesis version of declipping algorithm SPADE. Use the explicit projection formula, the computational cost, dominated by synthesis and analysis operators, is identical for both versions (per iteration). However, S-SPADE needs fewer iterations to converge, thus turning it to be significantly faster than A-SPADE. As a result, S-SPADE is preferable in real-time processing. On average, A-SPADE performs better in terms of $\Delta$SDR than S-SPADE does.

Experiments involving the parameters of the DGT/DFT show that the optimal window size differs for the algorithms. Whereas A-SPADE performs best with shorter windows, S-SPADE, on the contrary, prefers slightly longer windows. The influence of the window overlap is not negligible either—we have shown that the bigger the overlap is, the better the restoration results are obtained, in both algorithms.

Unfortunately, our results for S-SPADE differ from what the original paper [24] reports. The authors of [24] claim that S-SPADE performs slightly better than A-SPADE does in terms of $\Delta$SDR, and also that S-SPADE performs best with redundancy 4. Our results indicate quite the opposite; in particular, our S-SPADE performs worse in terms of $\Delta$SDR and performs best when redundancy is set to 1.

Our future work will be to investigate in greater depth the differences between the synthesis and the analysis model (and their influence on audio restoration methods). We also believe that introducing a psychoacoustic model could lead to higher declipping quality.

**Acknowledgement** The authors thank S. Kitić for providing us with his implementation of the SPADE algorithms and for discussion. The work was supported by the joint project of the FWF and the Czech Science Foundation (GAČR): numbers I 3067-N30 and 17-33798L, respectively. Research described in this paper was financed by the National Sustainability Program under grant LO1401. Infrastructure of the SIX Center was used.

# References

1. Abel, J., Smith, J.: Restoring a clipped signal. In: Acoustics, Speech, and Signal Processing, 1991. ICASSP-91., 1991 International Conference on. (Apr 1991) 1745–1748 vol.3
2. Adler, A., Emiya, V., Jafari, M., Elad, M., Gribonval, R., Plumbley, M.: A constrained matching pursuit approach to audio declipping. In: Acoustics, Speech and Signal Processing (ICASSP), 2011 IEEE International Conference on. (2011) 329–332
3. Bayram, I., Kamasak, M.: A simple prior for audio signals. IEEE Transactions on Acoustics Speech and Signal Processing **21**(6) (2013) 1190–1200
4. Bayram, I., Akykıldız, D.: Primal-dual algorithms for audio decomposition using mixed norms. Signal, Image and Video Processing **8**(1) (2014) 95–110
5. Bertsekas, D.: Nonlinear Programming. Athena Scientific, Belmont (1999)
6. Bilen, C., Ozerov, A., Perez, P.: Audio declipping via nonnegative matrix factorization. In: Applications of Signal Processing to Audio and Acoustics (WASPAA), 2015 IEEE Workshop on. (Oct 2015) 1–5
7. Boyd, S.P., Parikh, N., Chu, E., Peleato, B., Eckstein, J.: Distributed optimization and statistical learning via the alternating direction method of multipliers. Foundations and Trends in Machine Learning **3**(1) (2011) 1–122
8. Bruckstein, A.M., Donoho, D.L., Elad, M.: From sparse solutions of systems of equations to sparse modeling of signals and images. SIAM Review **51**(1) (2009) 34–81
9. Christensen, O.: Frames and Bases, An Introductory Course. Birkhäuser, Boston (2008)
10. Combettes, P., Pesquet, J.: Proximal splitting methods in signal processing. Fixed-Point Algorithms for Inverse Problems in Science and Engineering (2011) 185–212
11. Dahimene, A., Noureddine, M., Azrar, A.: A simple algorithm for the restoration of clipped speech signal. Informatica **32** (2008) 183–188
12. Defraene, B., Mansour, N., Hertogh, S.D., van Waterschoot, T., Diehl, M., Moonen, M.: Declipping of audio signals using perceptual compressed sensing. IEEE Transactions on Audio, Speech, and Language Processing **21**(12) (Dec 2013) 2627–2637
13. Donoho, D.L., Elad, M.: Optimally sparse representation in general (nonorthogonal) dictionaries via $\ell_1$ minimization. Proceedings of The National Academy of Sciences **100**(5) (2003) 2197–2202
14. Duhamel, P., Vetterli, M.: Fast fourier transforms: A tutorial review and a state of the art. Signal Processing **19** (1990) 259–299

15. Fong, W., Godsill, S.: Monte carlo smoothing for non-linearly distorted signals. In: 2001 IEEE International Conference on Acoustics, Speech, and Signal Processing. Proceedings (Cat. No.01CH37221). Volume 6. (2001) 3997–4000 vol.6
16. Gaultier, C., Bertin, N., Kitić, S., Gribonval, R.: A modeling and algorithmic framework for (non)social (co)sparse audio restoration. (11 2017)
17. Godsill, S.J., Wolfe, P.J., Fong, W.N.: Statistical model-based approaches to audio restoration and analysis. Journal of New Music Research **30**(4) (2001) 323–338
18. Gröchenig, K.: Foundations of time-frequency analysis. Birkhäuser (2001)
19. Harvilla, M.J., Stern, R.M.: Least squares signal declipping for robust speech recongnition (2014)
20. Holighaus, N., Wiesmeyr, C.: Construction of warped time-frequency representations on nonuniform frequency scales, part i: Frames. submitted, preprint available: arXiv:1409.7203 (2016)
21. Janssen, A.J.E.M., Veldhuis, R.N.J., Vries, L.B.: Adaptive interpolation of discrete-time signals that can be modeled as autoregressive processes. IEEE Trans. Acoustics, Speech and Signal Processing **34**(2) (4 1986) 317–330
22. Kitić, S., Jacques, L., Madhu, N., Hopwood, M., Spriet, A., De Vleeschouwer, C.: Consistent iterative hard thresholding for signal declipping. In: Acoustics, Speech and Signal Processing (ICASSP), 2013 IEEE International Conference on. (May 2013) 5939–5943
23. Kitić, S., Bertin, N., Gribonval, R.: Audio declipping by cosparse hard thresholding. In: 2nd Traveling Workshop on Interactions between Sparse models and Technology. (2014)
24. Kitić, S., Bertin, N., Gribonval, R.: Sparsity and cosparsity for audio declipping: a flexible non-convex approach. In: LVA/ICA 2015 – The 12th International Conference on Latent Variable Analysis and Signal Separation, Liberec, Czech Republic (August 2015)
25. Málek, J.: Blind compensation of memoryless nonlinear distortions in sparse signals. In: 21st European Signal Processing Conference (EUSIPCO 2013). (Sept 2013) 1–5
26. Miura, S., Nakajima, H., Miyabe, S., Makino, S., Yamada, T., Nakadai, K.: Restoration of clipped audio signal using recursive vector projection. In: TENCON 2011 - 2011 IEEE Region 10 Conference. (Nov 2011) 394–397
27. Nam, S., Davies, M., Elad, M., Gribonval, R.: The cosparse analysis model and algorithms. Applied and Computational Harmonic Analysis **34**(1) (2013) 30 – 56
28. Necciari, T., Balazs, P., Holighaus, N., Sondergaard, P.: The erblet transform: An auditory-based time-frequency representation with perfect reconstruction. In: Acoustics, Speech and Signal Processing (ICASSP), 2013 IEEE International Conference on. (May 2013) 498–502
29. Šorel, M., Bartoš, M.: Efficient jpeg decompression by the alternating direction method of multipliers. In: 2016 23rd International Conference on Pattern Recognition (ICPR). (Dec 2016) 271–276
30. Průša, Z., Søndergaard, P., Balazs, P., Holighaus, N.: LTFAT: A Matlab/Octave toolbox for sound processing. In: Proceedings of the 10th International Symposium on Computer Music Multidisciplinary Research (CMMR 2013), Marseille, France, Laboratoire de Mécanique et d'Acoustique, Publications of L.M.A. (October 2013) 299–314
31. Rajmic, P., Bartlová, H., Průša, Z., Holighaus, N.: Acceleration of audio inpainting by support restriction. In: 7th International Congress on Ultra Modern Telecommunications and Control Systems. (2015)

32. Selesnick, I.: Least squares with examples in signal processing (April 2013)
33. Siedenburg, K., Kowalski, M., Dorfler, M.: Audio declipping with social sparsity. In: Acoustics, Speech and Signal Processing (ICASSP), 2014 IEEE International Conference on, IEEE (2014) 1577–1581
34. Tachioka, Y., Narita, T., Ishii, J.: Speech recognition performance estimation for clipped speech based on objective measures. Acoustical Science and Technology **35**(6) (2014) 324–326
35. Tan, C.T., Moore, B.C.J., Zacharov, N.: The effect of nonlinear distortion on the perceived quality of music and speech signals. J. Audio Eng. Soc **51**(11) (2003) 1012–1031
36. Weinstein, A.J., Wakin, M.B.: Recovering a clipped signal in sparseland. CoRR **abs/1110.5063** (2011)